\documentstyle{article}

\title{Quantum Mechanics as a Classical Theory V: \\
The Quantum Schwartzchild Problem}
\author{L. S. F. Olavo\\
Departamento de Fisica - Universidade de Brasilia - UnB\\
70910-900 - Brasilia - D.F. - Brazil}

\begin{document}

\maketitle
\begin{abstract}
In this continuation paper, we apply the general relativistic quantum theory
for one particle systems, derived in paper II of this series, to a simple
problem: the quantum Schwartzchild problem, where one particle of mass {\it m%
} gravitates around a massive body. The results thus obtained reveal that,
in the realm of such a theory, the negative mass conjecture we made in paper
IV of this series is, indeed, adequate. It is shown that gravitation is
responsible for the loss of energy quantization. We relate this property
with the ideas of irreversibility and time arrow.
\end{abstract}

\section{Introduction}

In the previous paper (hereafter IV), we conjectured that it is possible to
describe Nature appealing to negative mass particles (antiparticles). All
the experiments the ''orthodox'' interpretation might explain can also be
explained by our interpretation. The experiment where particle and
antiparticle travel under the influence of a homogeneous magnetic field was
one we explained in details. We saw that the electromagnetic interaction
alone is not capable of deciding about mass sign, since it is concerned only
with the ratio $e/m$ and the velocity. We also argued that a theory that
accounts for gravitational effects will be capable of such a decision.

In paper II of this series\cite{1,2,3}, we developed a general relativistic
quantum theory for ensembles composed of one particle systems. This theory
can be considered an immediate generalization of Klein-Gordon's and Dirac's
special relativistic theories. It includes Einstein's equations as part of
the system of equations one shall solve and, thus, takes into account
gravitation. The application of this theory to a problem where only
gravitational effects are present will decide unambiguously about the
correctness of the conjecture made in IV.

In the present paper we will apply the above mentioned theory to what we
call the quantum Schwartzchild problem. The classical version of this
problem is well known: a massive body leads to a geometrical distortion in
space-time structure that is felt by a test particle with vanishing mass.

The quantum counterpart of this problem is as follows: we suppose that the
initial conditions related with the test particle might not be known. It is
thus necessary to approach the problem statistically. The resulting
statistical description shall account for the test particle probability
distribution over space-time. The function that emerges from the
calculations shall represent the probability amplitude related with the test
particle being somewhere in tridimensional space at some instant of time -
an event probability amplitude.

As with the electron clouds of the hydrogen atom problem, the test particle
becomes represented by a continuous (probability) density distribution. This
implies that all its properties, such as the mass or the charge, shall be
also considered as continuously distributed in space-time.

In the second section, we state the problem mathematically and solve it
exactly.

The third section will be concerned with the interpretations of the results
obtained in the second section.

We then make our conclusions. We discuss the effect of gravity upon
quantization and it is shown that gravitation is related with the extinction
of quantization. This property is paralleled with the idea of
irreversibility and time arrow. The superposition principle is also
discussed.

In the appendix, some of the more restringent classical arguments
against\cite{5,6} ''antigravity'' are discussed in details and it is shown
that they are not adequate.

\section{The Problem}

We showed, in paper II of this series, that the system of equations we shall
solve when considering a general relativistic quantum problem is given by
\begin{equation}
\label{1}\frac{-\hbar ^2}{2mR}\Box R+V-\frac{mc^2}2+\frac{\nabla _\beta
S\nabla ^\beta S}{2m}=0;
\end{equation}
\begin{equation}
\label{2}G_{\mu \nu }=-\frac{8\pi G}{c^2}\left[ T_{(M)\mu \nu }+T_{(Q)\mu
\nu }\right] ,
\end{equation}
where the functions $R$ and $S$ are related to the probability amplitude by
\begin{equation}
\label{3}\psi \left( x\right) =R\left( x\right) \exp \left( iS\left(
x\right) /\hbar \right) ,
\end{equation}
$G$ is Newton's gravitational constant, $T_{(M)\mu \nu }$ is the matter
energy-momentum tensor, $T_{(Q)\mu \nu }$ is the energy-momentum tensor of
the statistical field\cite{1,2,3} given, in terms of $R$ and $S$ as
\begin{equation}
\label{31}T_{(Q)\mu \nu }=mR\left( x\right) ^2\frac{\nabla _\mu S}m\frac{%
\nabla _\nu S}m
\end{equation}
and $G_{\mu \nu }$ is Einstein's tensor (here $m$ represents the modulus of
the mass appearing in the Klein-Gordon's equation).

In the problem in which we are interested there is only the gravitational
force. Then, we expect only the statistical field's energy-momentum tensor
to appear in the right side of equation (\ref{2}). Our specific problem
demands that we rewrite equations (\ref{1}) and (\ref{2}) as
\begin{equation}
\label{4}\frac{-\hbar ^2}{2mR}\Box R-\frac{mc^2}2+\frac{\nabla _\beta
S\nabla ^\beta S}{2m}=0
\end{equation}
and
\begin{equation}
\label{5}G_{\mu \nu }=-\frac{8\pi G}{c^2}\rho \left( r,\tau \right) u_\mu
u_\nu ,
\end{equation}
where we used the following conventions:
\begin{equation}
\label{6}\rho \left( x\right) =mR\left( x\right) ^2\ ;\ u_\mu =\frac{\nabla
_\mu S}m.
\end{equation}

It is preferable to treat this problem using comoving coordinates defined by
the line element
\begin{equation}
\label{7}ds^2=c^2d\tau ^2-e^{w\left( r,\tau \right) }dr^2-e^{v\left( r,\tau
\right) }\left( d\theta ^2+\sin {}^2\theta d\phi ^2\right) ,
\end{equation}
where $\tau $ is the particle proper time, $(r,\theta ,\phi )$ its
spherical-polar coordinates and $w\left( r,\tau \right) $, $v\left( r,\tau
\right) $ the functions we shall obtain to fix the metric. Looking at
equation (\ref{4}) we can see that, in the comoving coordinate system, we
shall have
\begin{equation}
\label{8}\nabla _\mu S\nabla ^\mu S=m^2c^2\Rightarrow u_\mu u^\mu =c^2,
\end{equation}
implying that
\begin{equation}
\label{9}S\left( x\right) =\pm mc^2\tau .
\end{equation}

As the coordinate system is comoving, we might put $u^\mu =(u^0,0,0,0)$ and
the statistical field's energy-momentum tensor becomes
\begin{equation}
\label{10}T_{(Q)00}=\rho \left( r,\tau \right) c^2\ ;\ T_{(Q)\mu \nu }=0\
\mbox{if}\ \mu \neq 0\ \mbox{or}\ \nu \neq 0.
\end{equation}
Einstein's equations can now be written explicitly as
\begin{equation}
\label{11}-e^{-w}\left( v^{\prime }{}^{\prime }+\frac 34v^{\prime 2}-\frac
12w^{\prime }v^{\prime }\right) +e^{-v}+\frac 14\stackrel{\cdot }{v}^2+\frac
12\stackrel{\cdot }{v}\stackrel{\cdot }{w}=8\pi G\rho ;
\end{equation}
\begin{equation}
\label{12}v^{\prime }+\frac 12w^{\prime }-\frac 12\stackrel{\cdot }{w}%
v^{\prime }=0;
\end{equation}
\begin{equation}
\label{13}e^w\left( \stackrel{\cdot \cdot }{v}+\frac 34\stackrel{\cdot }{v}%
^2+e^{-v}\right) -\frac 14v^{\prime 2}=0;
\end{equation}
\begin{equation}
\label{14}e^v\left( \stackrel{\cdot \cdot }{v}+\frac 14\stackrel{\cdot }{v}%
^2+\frac 14\stackrel{\cdot }{v}\stackrel{\cdot }{w}+\frac 12\stackrel{\cdot
\cdot }{w}+\frac 14\stackrel{\cdot }{w}^2\right) +e^{v-w}\left( \frac
14w^{\prime }v^{\prime }-\frac 12v^{\prime }{}^{\prime }-\frac 12v^{\prime
2}\right) =0,
\end{equation}
where the line and the dot indicate derivatives regarding variables $r$ and $%
\tau $, respectively. We can solve the last three equations if we put\cite{4}
\begin{equation}
\label{15}e^w=\frac{e^vv^{\prime 2}}4\ ;\ e^v=\left[ F\left( r\right) \tau
+G\left( r\right) \right] ,
\end{equation}
where $F(r)$ and $G(r)$ are arbitrary functions of $r$. From equation (\ref
{11}) we get the density function $\rho \left( r,t\right) $ with its
explicit dependence on the metric given by the functions $F(r)$ and $G(r)$:
\begin{equation}
\label{16}\rho \left( r,\tau \right) =\left[ \frac 1{6\pi G}\right] \frac{%
F\left( r\right) F^{\prime }\left( r\right) }{\left[ F\left( r\right) \tau
+G\left( r\right) \right] \left[ F^{\prime }\left( r\right) \tau +G^{\prime
}\left( r\right) \right] }.
\end{equation}

To solve our primary system of equations (\ref{1}-\ref{2}) we still have to
solve equation (\ref{1}) that becomes, using (\ref{8}) and (\ref{9}),
\begin{equation}
\label{17}\Box R\left( r,\tau \right) =0.
\end{equation}
We shall stress at this point that equation (\ref{17}) is highly non-linear.
The functions that define the density also define the metric. These
functions will equally well be present in the D'Alambertian operator.
Moreover, the function $R(r,\tau )$ is the square-root of the density
function given by (\ref{16}).

We can solve this equation using the degree of freedom we have in the choice
of the arbitrary function $G(r)$; choosing it to be identically zero
\begin{equation}
\label{18}G\left( r\right) =0,
\end{equation}
we obtain the result
\begin{equation}
\label{19}R\left( r,\tau \right) =N\sqrt{\frac 1{6\pi mG}}\frac 1\tau ,
\end{equation}
where $N$ is a normalization constant.

Replacing these results in the expression (\ref{7}) for the metric, we get
\begin{equation}
\label{20}ds^2=c^2d\tau ^2-\left( \frac 49\frac{F^{\prime }\left( r\right)
^2\tau ^2}{F\left( r\right) ^{2/3}\tau ^{2/3}}\right) dr^2-\left[ F\left(
r\right) \tau \right] ^{4/3}\left( d\theta ^2+\sin {}^2\theta d\phi
^2\right) ,
\end{equation}
that can be further reduced to the format
\begin{equation}
\label{21}ds^2=c^2d\tau ^2-\tau ^{4/3}\left[ d\chi ^2-\chi ^2\left( d\theta
^2+\sin {}^2\theta d\phi ^2\right) \right] ,
\end{equation}
where
\begin{equation}
\label{22}\chi \left( r\right) =F\left( r\right) ^{2/3}.
\end{equation}

Collecting all the above results, the probability amplitude for the quantum
Schwartzchild problem becomes
\begin{equation}
\label{23}\psi _P\left( \tau \right) =N\sqrt{\frac 1{6\pi mG}}\frac{%
e^{-imc^2\tau /\hbar }}\tau \ ;\ \psi _A\left( \tau \right) =N\sqrt{\frac
1{6\pi mG}}\frac{e^{+imc^2\tau /\hbar }}\tau
\end{equation}
in the comoving coordinate system, representing particle and antiparticle
solutions (here we are supposing the massive body to be made of positive
mass but this is not crucial; the important thing here is the difference in
the signs of the probability amplitudes phases).

The interpretation of equation (\ref{23}) is unambiguous. The particle
solution represents the probability density that a particle, in its rest
frame, is traveling in the direction of the massive body along its own
world-line. The solution represented by $\psi _A$ gives a particle that is
traveling along its world-line, but in the contrary proper time direction
(figure 1a). We might draw a better picture of what is happening if we take
a look at the projection of these world trajectories into three dimensional
space. It becomes clear that, while the particle is falling freely over the
massive body, the antiparticle moves farther in what we might call free
ejection.

To adequate this to a description where proper time flows only in the
positive direction, it is necessary that we invert the time coordinate of
the plus sign solution. This will make the trajectory of the plus sign
solution coincide with the other particle trajectory. To achieve again the
free ejection in three dimensional space we shall also invert the sign of
the mass (figure 1b). This is precisely the same procedure already done in
the appendix of the previous paper.

We stressed, in the last paper, that these interpretations are
mathematically, though not physically, equivalent. This theory thus presents
a symmetry; it says that a positive mass particle flowing backward in time
is equivalent to a negative mass antiparticle flowing onward. It then turns
out that particles are attracted by the gravitational field of a positive
mass body, while antiparticles are repelled. This fixes the signs of masses.

The resulting metric is of a Robertson-Walker type
\begin{equation}
\label{24}ds^2=c^2d\tau ^2-R\left( \tau \right) \left[ \frac{dr^2}{1-kr^2}%
+r^2\left( d\theta ^2+\sin {}^2\theta d\phi ^2\right) \right] ,
\end{equation}
with $k=0$; meaning that three space, with the radius defined by (\ref{22}),
is flat. The Hubble constant is easily computed and gives the usual value
\begin{equation}
\label{25}H=\frac 1\tau ,
\end{equation}
as expected for this problem.

\section{Conclusions}

We might derive many interesting consequences from the previous formalism
and the example above.

The first thing we note when looking at system (\ref{1}-\ref{2}) is that
this system is highly non-linear. This feature can be exemplified by
expression (\ref{17}) of the quantum Schwartzchild problem. We do not have,
in general, a linear eigenvalue equation.

Energy quantization is, however, strictly related with the quantum equations
being linear eigenvalue ones. If a system is to be described by the system
of equations (\ref{1}-\ref{2}), then we shall not expect energy quantization
to take place. Although some systems might still keep, in some very
restricted range, their property of quantization, this shall be the
exception rather than the rule.

The tendency of the energy spectrum to lose its discreteness character might
be associated with a departure from the equilibrium by the system. Indeed,
in our previous calculations\cite{1,2,3}, we started with the classical
Liouville's equation and then obtained the quantum equation for the
probability density. This density was then written as
\begin{equation}
\label{26}\rho \left( x,x^{\prime }\right) =\psi ^{\dagger }\left( x^{\prime
}\right) \psi \left( x\right) ,
\end{equation}
and we were able to derive Schr\"odinger's equation (non-relativistic case)
for the probability amplitude. When we assume a stationary, pure state,
configuration for the system
\begin{equation}
\label{27}\psi \left( {\bf x},t\right) =\varphi \left( {\bf r}\right)
e^{-iEt/\hbar },
\end{equation}
we are also assuming that the probability density does not depend explicitly
on the time. This is equivalent to assume that the system is in one of its
equilibrium states.

With the generalization introduced by system (\ref{1}-\ref{2}), it is, in
general, not possible to admit a time dependence like (\ref{27}) above. The
superposition principle shall not be valid since it is based in the linear
character of the equations. The gravitational field, thus, plays the role of
removing the system from equilibrium.

This might be seen in the example of the quantum Schwartzchild problem. If
we are to solve this problem, non relativistically, in flat space, we shall
use Schr\"odinger's equation with the gravitational potential
\begin{equation}
\label{28}V=-\frac{GM}r.
\end{equation}
We then get a level scheme, basically a hydrogen-like one, where the levels
will be very nearly spaced, but we still get quantization. When the problem
is solved, using system (\ref{1}-\ref{2}), all the quantization disappears
and we are left with a collapse-ejection-like solution. We stress that this
feature is expected for almost all the problems (note that we also do not
have solutions depending on the angles). We might thus say that we are lucky
in living in a world where gravitation does not play a predominant role. Our
world might be thought as a world in equilibrium only to a good
approximation, but not strictly.

Gravity pushes everything to non-equilibrium states until, probably,
col\-la\-pse-e\-jec\-tion takes place. In strong gravitational fields we shall
not expect to meet atoms as we face them in our everyday life. They shall
not be in their stable configurations.

When driving all systems to non-equilibrium configurations, gravitation
introduces a time arrow . In the example of the atom cited above, even if
the atom (hydrogen, for simplicity) is distant from any other massive body,
it is suficient that the proton and the electron have finite masses for
introducing into the system a finite mean lifetime. Of course, since their
gravitational field is very feeble, the system's mean lifetime will be
enormous. We might also consider that electromagnetic forces are also
present, implying that the solution obtained in the last section will not
strictly apply. However, we do not expect the qualitative analysis to be
much different. We might, of course, find in Nature other forces capable of
stabilizing a system. These forces, however, will compete with gravitation.

We thus began trying to consolidate the negative mass conjecture and arrived
at a striking different world. We might see that the world suggested by
these results is very distinct from the one we are familiarized when we
consider the questions about the cosmological models of our Universe and
compare their traditional answers with the ones emerging from our picture of
Nature. This will be left for another work.

It is important to stress here that, for the epistemological framework
adopted by this series of papers, the words (and worlds) {\it classical} and
{\it quantum} are not oposed to each other; they are, on the contrary,
complemetary views of the same (classical) Nature. Quantum Mechanics is here
merely a name for a classical statistical mechanics (from the ontological
point of view) performed in configuration space. Indeed, for the orthodox
quantum mechanical view, the notion of geodesic is not admissible\cite{6}.
This theory, therefore, does not suffer from that sort of
''incompatibilities'' emphasized by a great number of authors\cite{7}-
\cite{11}. That is why it was possible to join them into just one theory
without modifying their structures\cite{12,13}.

One last word is, nevertheless, appropriate. This theory might explain why
our universe seems to have many more particles than antiparticles. The
property of gravitation to be an attractive (long range) force between
similar entities while being a repulsive force between different ones
implies that the Universe tends to split into two distinct parts\cite{14}.
Considering that some radiation era existed, when pairs of particle and
antiparticle were being created, their mutual repulsion might have strongly
impeled them to occupy distinct portions of the Universe in a cumulative
process\cite{15}.

It is extremely interesting to see how two highly different worlds emerged
from two distinct interpretations of the special relativistic formalism of
quantum mechanics.

There have appeared in the literature since 1957 many arguments against the
notion of ''antigravity''\cite{5,6,16,17}. Many of them are based in
gedanken experiments and, as we have shown in the appendix, cannot be
sustained. Some experiments are now in progress to measure the gravitational
acceleration of antiparticles\cite{6}. They will be of utmost importance to
prove, or disprove, the negative mass conjecture.

\appendix

\section{Antigravity}

The first ideas of ''antigravity'' came into play in 1957 with the
pioneering work of Bondi\cite{18} who asked if general relativity could
accomodate the notion of negative mass. Yet, in the following year, the
first argument against negative masses appeared with Morrison's celebrated
paper\cite{5,19}. Other arguments followed in the subsequent years\cite{%
16,17}. Morrison's argument, however, might be considered the most
restricting one, since it is related with the notion of energy conservation
and is the one we will consider in more detail in this appendix.

Morrison's argument is based on a {\it gedanken} experiment. Thus, before
analyzing the argument itself, it is interesting to say some words about the
{\it status} we shall ascribe to such an approach.

{\it Gedanken} experiments are not, as one may think, a peculiarity of our
century. Indeed, it comes from the aristotelian idea that one might
scrutinize Nature by means of thought alone and is based on the assumed
homogeneity of human's reasoning and Nature's order\cite{20}. This idea,
however, does not take into account that, when performing any {\it gedanken}
experiment, we are also approaching Nature by means of some theory (or
proto-theory). This is nothing but the dispute between Hume\cite{21} and
Kant\cite{22}, to say but a few. This can be verified by historical examples.

The first one we cite is the debate between aristotelians and galileans
about the free fall of bodies. Both schools based their arguments on the
same mental experiment\cite{20}, the projectile falling from the mast of a
ship in movement. Their conclusions were, however, opposite. The
aristotelians, based on their conceptions of Nature, should not accept that
the {\it impetus} (movement, in modern terms) of the ship was transferred to
the projectile, and so, once left it will never fall at the base of the
mast. The galileans, however, had a picture of Nature that could accomodate
the {\it impetus} transference; for them the projectil will ever fall at the
base of the mast. Although we now know the galileans were correct, none of
them has ever done the experiment\cite{20} at that time. It was a question
of principle.

The other famous series of {\it gedanken} experiments are those related with
the dispute between Einstein and Bohr. One of them, the EPR\cite{23},
playing a relevant role on the epistemological development of quantum
mechanics. Looking at them carefully might convince one that both sides are
full of epistemological prejudices\cite{24,25}.

The considerations made above do not intend to deny the importance of {\it %
gedanken} experiments (some times they are the only ones left). They have
just the intention of clarifying that these experiments shall not be used to
exclude Nature behavior but only exclude Nature behavior {\it with respect
to some theory or approximation}.

The arguments above might be considered a runaway solution. To avoid this
misinterpretation we will present Morrison's arguments and show that they
can be used also to get the opposite answer.

\subsection{Morrison's Argument:}

Morrison bases his arguments about negative mass beginning with an
experiment with positive masses and then extending it to embrace negative
ones. He then shows that this leads to a violation of the energy
conservation principle. We begin, thus, with this first experiment where
only positive masses are considered.

The experimental setup is shown in figure 2. It consists of a well-balanced
and friction-free Atwood's machine mounted in a uniform time-constant
gravitational field. The axle of the upper pulley is belted to an energy
storing device at the upper level marked simply ''output''. With no leads,
the dumbwaiter moves freely and without energy loss or gain. With the
appropriate clock setting, Morrison's argument is as follows:

''Place an atom in the lower dumbwaiter pan, at gravitational potential $%
\phi _1$, and an identical atom, but excited to its first quantum level
above ground, in the second pan at gravitational potential $\phi _2$. The
upper pan is then heavy by the weight of its extra energy content, $%
mg=g\Delta E/c^2$. The heavier pan will fall. We can allow it to reach a
small velocity, and then keep it from accelerating further by drawing an
output current from the coupled generator, storing the energy in the storage
cells. When the pans have exchanged places, I let the excited atom down
below to decay to its ground state. In the successful trials, the photon
emitted comes up to the other pan, where it strikes the upper atom, in its
ground state. Were the photon to excite the upper atom to the first excited
level, I should have restored the initial condition, and yet have collected
energy in the storage cell. I must avoid this by the hypotesis of the first
law [energy conservation]. This I can do if I realize that the photon is red
shifted, having insufficient energy to excite the atom by just the amount
stored in the cell.''

With this reasonings, Morrison claims that the red-shift formula can be
derived using only the equivalence between gravitational and inertial mass
and the need to obey the overall conservation of energy.

There are, however, two major problems with these reasonings: first, they do
not take into account that the energy levels of the atoms should be
distorted by the gravitational field. Second, the isolated system is
comprised by the atoms plus the gravitational field. Indeed, even the idea
of energy in the realm of Einstein's gravitation theory is not as clear as
in Newton's. Despite this last consideration, let us analyse Morrison's
argument and see if it can be reformulated.

Consider now the arrangement of figure 3 where we show the same apparatus
with the atoms energy levels on its right side. The atom1 has its first
excited level lowered by the gravitational field by an amount of $\delta E$
with respect to atom2. Due to its greater energy content, $mg=g(\Delta
E-\delta E)/c^2$, atom1 falls down freely. As atom1 arrives at the bottom,
its energy content is now $mg=g\Delta E/c^2$, since the gravitational field
has changed with height. The gravitational field has lost the energy amount $%
\delta E$ realizing work on the levels of atom1. Now this configuration
implies that the difference between the ground and excited levels of atom2
is $(\Delta E-\delta E)$. Let atom1 decay. We might build the Atwood's
machine to have a height such that the red-shift of the emitted photon is
exactly $\delta E$. The photon thus have the exactly amount of energy to
excite atom2 and the process, thus, continues. The system
atoms-plus-gravitational-field is isolated and energy was conserved since
the amount of energy given by the field to atom1 was restored when the field
extracted energy from the photon by virtue of the red-shift. If Atwood's
machine is not calibrated as above, the photon will not be capable of
exciting atom2 and the process will stop. In this case, we still have energy
conservation when considering the system
atoms-plus-gravitational-field-plus-photon (red-shifted).

The lesson to learn is that energy conservation is not due to the photon
red-shift. Indeed, in Morrison's original experiment (with energy drain {\it %
and} the distortion of levels) it is possible to calibrate Atwood's machine
in such a way that the photon still has sufficient energy to excite atom2.
In this case, energy was taken from the gravitational field reducing its
mass (in a process we might call evaporation). Since the mass responsible
for the gravitational field is finite, this process of energy extraction
from the gravitational field must stop (thus, in the formulae above, when
energy drain is considered, we should also take into account the change $%
\delta g$ of gravitational acceleration by virtue of mass loss). We also do
not expect this process of energy extraction to have high efficience since
we are supposing that we can control the time when the atom decays,
something we cannot do.

In the second experiment, Morrison introduces the concept of negative
mass. For this experiment we will use a more elaborated version\cite{6}
having the same physical content. The experiment is as follows:

Take a particle-antiparticle pair with equal positive and negative masses
respectively and place it at rest on a gravitational field. Since the pair
has no net weight it might be suspended adiabaticaly to some height $L$.
Then, let the pair annihilate. The produced pair of photons will travel back
to height zero suffering a blue-shift. Now, at the botton, let the photons
produce the pair again. Since the photon energy is now greater, because of
the blue-shift, the pair will have some extra kinetic energy with respect to
the beginning of the process. The conclusion is that energy was created\cite{%
5} violating the energy conservation principle.

The problem here is the same. The pair is considered isolated. However, the
isolated system is the pair-plus-gravitational-field. In this case, the
extra kinetic energy gained by the pair is exactly the amount extracted from
the gravitational field with the photon's blue-shift. When considering the
system pair-plus-gravitational-field, energy is conserved in exactly the
same way as above. The process is just an inventive way, although not too
efficient, of extracting energy from the gravitational field.

The fail of the energy-conservation argument can not be accepted.

\newpage

\unitlength=1.00mm
\special{em:linewidth 1pt}
\linethickness{1pt}

\begin{figure}
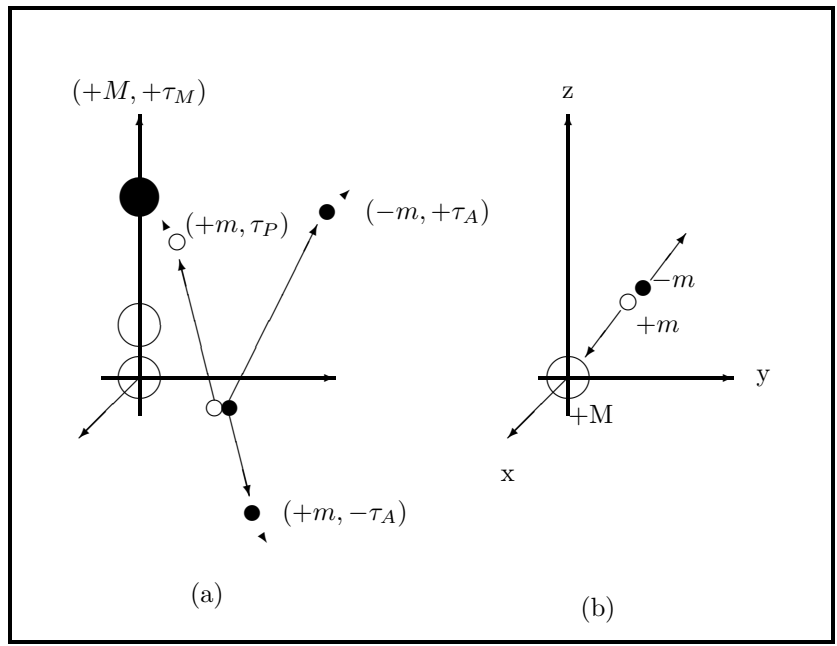
(123.00,84.00)
\put(31.00,30.00){\vector(0,1){40.00}}
\put(26.00,35.00){\vector(1,0){31.00}}
\put(31.00,35.00){\vector(-1,-1){8.00}}
\put(31.00,59.00){\circle*{5.20}}
\put(31.00,35.00){\circle{6.00}}
\put(31.00,42.00){\circle{6.32}}
\put(41.00,31.00){\circle{2.00}}
\put(43.00,31.00){\circle*{2.00}}
\put(50.00,17.00){\makebox(0,0)[lc]{$(+m,-\tau_{A})$}}
\put(61.00,57.00){\makebox(0,0)[lc]{$(-m,+\tau_{A})$}}
\put(56.00,57.00){\circle*{2.00}}
\put(57.00,58.00){\vector(1,1){2.00}}
\put(46.00,17.00){\circle*{2.00}}
\put(46.00,16.00){\vector(2,-3){2.00}}
\put(31.00,73.00){\makebox(0,0)[cc]{$(+M,+\tau_{M})$}}
\put(36.00,53.00){\circle{2.00}}
\put(35.00,54.00){\vector(-1,2){1.00}}
\put(37.00,57.00){\makebox(0,0)[lt]{$(+m,\tau_{P})$}}
\put(40.00,6.00){\makebox(0,0)[cc]{(a)}}
\put(88.00,30.00){\vector(0,1){40.00}}
\put(84.00,35.00){\vector(1,0){26.00}}
\put(88.00,35.00){\vector(-1,-1){8.00}}
\put(88.00,35.00){\circle{6.00}}
\put(96.00,45.00){\circle{2.00}}
\put(98.00,47.00){\circle*{2.00}}
\put(95.00,44.00){\vector(-3,-4){4.67}}
\put(99.00,48.00){\vector(3,4){4.67}}
\put(80.00,22.00){\makebox(0,0)[cc]{x}}
\put(114.00,35.00){\makebox(0,0)[cc]{y}}
\put(88.00,73.00){\makebox(0,0)[cc]{z}}
\put(97.00,42.00){\makebox(0,0)[lc]{$+m$}}
\put(102.00,48.00){\makebox(0,0)[cc]{$-m$}}
\put(92.00,6.00){\makebox(0,0)[ct]{(b)}}
\put(41.00,32.00){\vector(-1,4){4.67}}
\put(43.00,30.00){\vector(1,-4){2.67}}
\put(43.00,32.00){\vector(1,2){11.67}}
\put(91.00,30.00){\makebox(0,0)[cc]{+M}}
\put(14.00,0.00){\framebox(109.00,84.00)[cc]{}}
\caption{Behavior of a particle-antiparticle pair in the presence
of a gravitational field generated by a body of mass +M. (a) in the
four-space (b) in the three-space. Particle is represented by empty circle
and antiparticle by filled circle.}
\end{figure}

\newpage
\begin{figure}
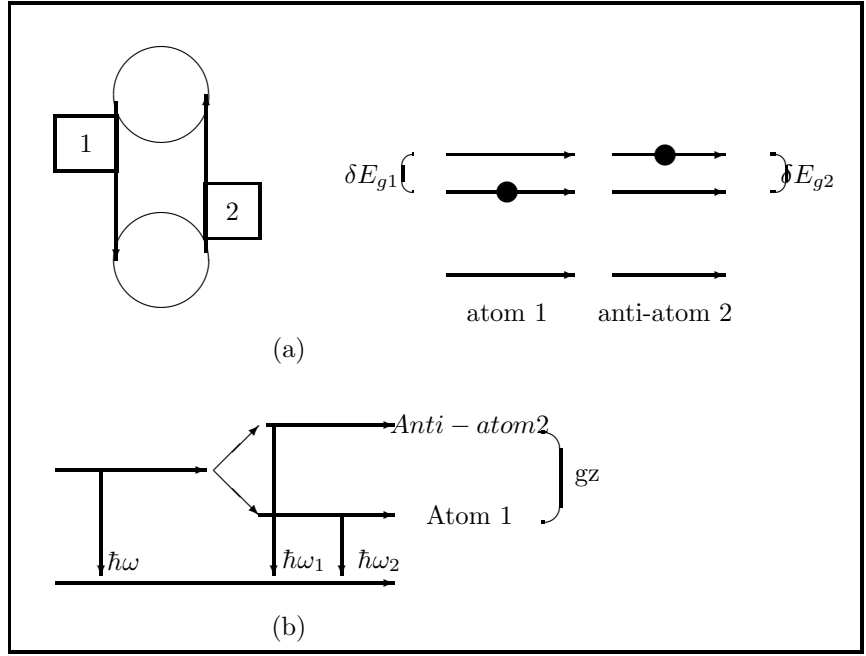
(116.00,87.00)
\put(9.00,10.00){\vector(1,0){45.00}}
\put(9.00,25.00){\vector(1,0){20.00}}
\put(30.00,25.00){\vector(1,1){6.00}}
\put(37.00,31.00){\vector(1,0){17.00}}
\put(30.00,25.00){\vector(1,-1){6.00}}
\put(36.00,19.00){\vector(1,0){18.00}}
\put(15.00,25.00){\vector(0,-1){14.00}}
\put(38.00,31.00){\vector(0,-1){20.00}}
\put(47.00,19.00){\vector(0,-1){8.00}}
\put(18.00,13.00){\makebox(0,0)[cc]{$\hbar\omega$}}
\put(42.00,13.00){\makebox(0,0)[cc]{$\hbar\omega_1$}}
\put(52.00,13.00){\makebox(0,0)[cc]{$\hbar\omega_2$}}
\put(64.00,19.00){\makebox(0,0)[cc]{Atom 1}}
\put(64.00,31.00){\makebox(0,0)[cc]{$Anti-atom 2$}}
\put(73.50,24.00){\oval(5.00,12.00)[r]}
\put(78.00,24.00){\makebox(0,0)[lc]{gz}}
\put(40.00,4.00){\makebox(0,0)[cc]{(b)}}
\put(23.00,53.00){\circle{12.17}}
\put(23.00,75.00){\circle{12.00}}
\put(29.00,54.00){\vector(0,1){21.00}}
\put(17.00,74.00){\vector(0,-1){21.00}}
\put(29.00,56.00){\framebox(7.00,7.00)[cc]{2}}
\put(9.00,65.00){\framebox(8.00,7.00)[cc]{1}}
\put(40.00,41.00){\makebox(0,0)[cc]{(a)}}
\put(61.00,51.00){\vector(1,0){17.00}}
\put(83.00,51.00){\vector(1,0){15.00}}
\put(61.00,67.00){\vector(1,0){17.00}}
\put(83.00,67.00){\vector(1,0){15.00}}
\put(61.00,62.00){\vector(1,0){17.00}}
\put(83.00,62.00){\vector(1,0){15.00}}
\put(69.00,62.00){\circle*{2.83}}
\put(90.00,67.00){\circle*{2.83}}
\put(104.00,64.50){\oval(4.00,5.00)[r]}
\put(109.00,64.00){\makebox(0,0)[cc]{$\delta E_{g2}$}}
\put(56.50,64.50){\oval(3.00,5.00)[l]}
\put(51.00,64.00){\makebox(0,0)[cc]{$\delta E_{g1}$}}
\put(69.00,46.00){\makebox(0,0)[cc]{atom 1}}
\put(90.00,46.00){\makebox(0,0)[cc]{anti-atom 2}}
\put(3.00,1.00){\framebox(113.00,86.00)[cc]{}}
\caption{The matter-antimatter experiment. (a) an Atwood machine with two
pans. The higher contains matter while the lower antimatter. The atomic
levels of the atom and anti-atom are also represented. (b) The level diagram
of atoms and anti-atoms in the approximate gravitational potential $gz$.}
\end{figure}

\newpage

\begin{figure}
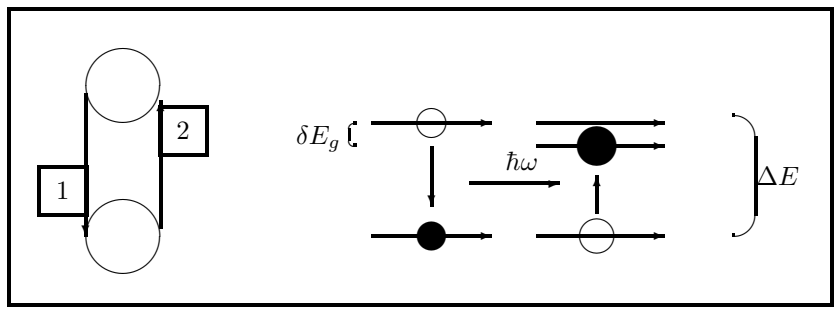
(113.00,42.00)
\put(19.00,32.00){\circle{10.20}}
\put(19.00,12.00){\circle{10.00}}
\put(14.00,31.00){\vector(0,-1){19.00}}
\put(24.00,13.00){\vector(0,1){17.00}}
\put(24.00,23.00){\framebox(6.00,6.00)[cc]{2}}
\put(8.00,15.00){\framebox(6.00,6.00)[cc]{1}}
\put(52.00,27.00){\vector(1,0){16.00}}
\put(52.00,12.00){\vector(1,0){16.00}}
\put(74.00,12.00){\vector(1,0){17.00}}
\put(74.00,27.00){\vector(1,0){17.00}}
\put(60.00,27.00){\circle{4.00}}
\put(60.00,12.00){\circle*{4.00}}
\put(60.00,24.00){\vector(0,-1){8.00}}
\put(65.00,19.00){\vector(1,0){12.00}}
\put(72.00,22.00){\makebox(0,0)[cc]{$\hbar\omega$}}
\put(74.00,24.00){\vector(1,0){17.00}}
\put(82.00,12.00){\circle{4.47}}
\put(82.00,24.00){\circle*{5.20}}
\put(82.00,15.00){\vector(0,1){5.00}}
\put(100.00,20.00){\oval(6.00,16.00)[r]}
\put(106.00,20.00){\makebox(0,0)[cc]{$\Delta E$}}
\put(50.00,25.50){\oval(2.00,3.00)[l]}
\put(45.00,25.00){\makebox(0,0)[cc]{$\delta E_g$}}
\put(4.00,3.00){\framebox(109.00,39.00)[cc]{}}

\caption{The field takes from the photon the amount of energy it gave to
the first atom when accelerating downward. This amount is exactly the
difference of energy levels of the second atom now placed at the top of
the Atwood machine.}
\end{figure}

\end{document}